# A Note on the Nonlinear Attenuation of an Ultrasound Contrast Agent Calculated from Rayleigh-Plesset Equation

Lang Xia

Email: lang4xia@gmail.com

**Abstract**

The attenuation of small-amplitude acoustic waves in a suspension containing ultrasound contrast agents (UCAs, coated microbubbles) is determined by the linear oscillation of the UCAs in the medium, which can be estimated via a linear attenuation theory. Recently, several nonlinear phenomena of energy attenuation at very low-intensity of acoustic pressures have been observed experimentally, raising concerns on the validity of the linear attenuation theory. Explanations of the nonlinear phenomenon are still lacking. Particularly, the interpretation of the pressure-dependent attenuation phenomenon is still under debate. In this note, we investigated the energy dissipation of a single UCA via a nonlinear Rayleigh-Plesset equation and used a formula capable of estimating attenuation coefficient due to the nonlinear oscillation of the UCA. The simulation results show the linear oscillation of an UCA at low excitation pressures does not always guarantee the linearity in the energy attenuation. Although nonlinear oscillation of the UCA contributes to the occurrence of nonlinear attenuation phenomena, it is not the only trigger.

## I. INTRODUCTION

Microbubbles, coated with lipids, proteins, and polymers, have been applied in the medical field as Ultrasound Contrast Agents (UCAs) for diagnostic and therapeutic purposes due to their rich dynamic behaviors under the excitation of ultrasound (de Jong et al., 2000; Doinikov, Haac, & Dayton, 2009; Hoff, Sontum, & Hovem, 2000; Karandish et al., 2018; Kulkarni et al., 2018; Xia et al., 2017). The dynamic behaviors of UCAs in a liquid--governed by various types of Rayleigh-Plesset equations--contribute mostly to the attenuation of propagating acoustic waves in the suspension. A linear attenuation theory that is derived from linearized Rayleigh-Plesset equations is usually employed to determine the attenuation coefficient of the suspension of UCAs (Church, 1995; Hoff et al., 2000). In this theory, a microbubble is analog to a driven harmonic oscillator, of which the energy dissipation at the linear region is calculated. The acoustic attenuation coefficient is then obtained by the summation the energy loss of each bubble (extinction cross-section) in the unit volume (Medwin, 1977).

Recent experimental observations (Gong, Cabodi, & Porter, 2014; Xia et al., 2014; Xia, Porter, & Sarkar, 2015) on UCAs with lipid shells have shown nonlinear behaviors of the attenuation--such as pressure-dependent attenuation and the shift of the main resonance of an attenuation curve-- even when the excitation pressure was sufficient low. At such the low excitation pressure, the UCA is usually assumed to oscillate linearly. The attenuation coefficient estimated by the linear attenuation theory is thus independent of the excitation pressures. Therefore, the linear attenuation theory is incapable of interpreting the nonlinear attenuation phenomena. While the nonlinear attenuation is not likely caused by the nonlinear propagation of the acoustic waves as the bubbly medium is highly dispersive (Xia, 2019), a nonlinear shell theory was proposed and successfully explained the pressure-dependent attenuation that was observed in studying lipid-coated UCAs (Xia et al., 2015). Nevertheless, this interpretation may be not readily applied to polymer-coated UCAs that also showed nonlinear attenuation phenomenon in the experimental characterization (Xia, 2018; Xia et al., 2014). Therefore, it is worthwhile revisiting the origin of the nonlinear attenuation by checking the dynamic behavior of an UCA.

In this note, we numerically investigate the dynamics of a single UCA at low excitation pressures where its oscillation is usually assumed linear. By the employment of a formula capable of calculating attenuation due to the nonlinear oscillation of the UCA, we study the variation of attenuation of the UCA with respect to the excitation frequency pressures. Finally, we discuss the possible causes of the nonlinear attenuation phenomena.

## II. LINEAR ATTENUATION THEORY

The Rayleigh-Plesset equation describing the dynamics of a spherical microbubble can be written as

$$\rho\left(R\ddot{R}+\frac{3}{2}\dot{R}^2\right)+\frac{2\gamma}{R}+4\mu\frac{\dot{R}}{R}=p_g-p_\infty \quad (1)$$

where $\rho$ is the density of the surrounding liquid, $R$ is the instantaneous radius of the bubble and $\dot{R}=\partial R/\partial t$, $\ddot{R}=\partial^2 R/\partial t^2$, $\gamma$ is the surface tension, $\mu$ is the viscosity of the host liquid, $p_g$ is the pressure inside the bubble, and $p_\infty$ is the ambient pressure in the liquid. An UCA is a microbubble encapsulated by a shell material. Various materials, such as lipids, polymers, and proteins, can be used to encapsulate a free bubble, giving rise to different types of Rayleigh-Plesset equations. Since the shell material is not the focus for the present study, we simply assume the shell of an UCA to be viscoelastic (Sarkar, Shi, Chatterjee, & Forsberg, 2005). The dynamical equation for the UCA of equilibrium radius $R_0$, undergoing forced linear spherical pulsations [$R(t)=R_0+X(t)$ and $|X(t)|<<R_0$] at an external excitation pressure can be written in the form of

$$\ddot{X} + \omega_0 \delta \dot{X} + \omega_0^2 X = F\cos(\omega t) \qquad (2)$$

with

$$\omega_0^2 = \frac{1}{\rho R_0^2}(3\kappa P_0 - \frac{4\gamma_0}{R_0} + \frac{4E^s}{R_0}) \qquad (3)$$

$$\delta_l = \frac{4\mu}{\omega_0 R_0^2 \rho}, \quad \delta_s = \frac{4\kappa^s}{\omega_0 R_0^3 \rho}, \quad F = \frac{P_A}{\rho R_0} \qquad (4)$$

where $X$ is the displacement around equilibrium radius $R_0$, $\omega_0$ is the bubble's pulsation angular resonance frequency, $\kappa$ is the polytropic constant, $P_0$ is the amplitude of the ambient pressure, $\gamma_0$ is a reference value of the interfacial tension. $\delta = \delta_l + \delta_s$ is the damping constant, of which the terms in the right-hand side stand for dampings of liquid viscosity and interface, respectively, $\kappa^s$ and $E^s$ are the dilatational viscosity and elasticity of the bubble shell, respectively, $P_A$ is the amplitude of the excitation pressure. Eq.(2) indicates that the linear dynamics of a coated microbubble (UCA) is a linear harmonic oscillator, having a steady-state solution in the form of

$$X = \frac{F}{\left[(\omega_0^2 - \omega^2)^2 + (\delta\omega_0\omega)^2\right]^{1/2}} \cos(\omega t - \phi)$$

$$\phi = \tan^{-1}\left(\frac{\delta\omega_0\omega}{\omega_0^2 - \omega^2}\right) \qquad (5)$$

To estimate the oscillation induced attenuation, we can investigate the energy absorption of a bubble in the acoustic field (Xia, 2018). By assuming that an UCA oscillates spherically without thermal dissipation in an incompressible liquid, the energy delivered to the UCA can be written as

$$\Pi_i = (4\pi R^2 p)\dot{R} \qquad (6)$$

where $p = P_A \cos(\omega t)$ is the driven pressure. The average power delivered into the system is

$$\Pi = \frac{1}{T}\int_0^T \Pi_i dt \qquad (7)$$

where $T = 2\pi/\omega$ is the period. The average intensity of the impinging acoustic wave can be written as $I = P_A^2/(2\rho c_0)$. Here $c_0$ is the sound speed in the host liquid, and the extinction cross-section is given by

$$\sigma_e = \frac{\Pi}{I} = \frac{4\omega\rho c_0}{P_A}\int_0^{2\pi/\omega} R^2 \dot{R} \cos(\omega t) dt \qquad (8)$$

The above equation is the extinction cross-section of a microbubble. Here we only investigate the effects of bubble oscillation on the energy attenuation, detail discussions on other mechanisms of dissipation may refer to a comprehensive review (Ainslie & Leighton, 2011). By substituting the steady state solution Eq.(5) into Eq.(8), a linear attenuation theory in terms of the extinction cross-section $\sigma_e$ is given by the following equation (Medwin, 1977; Xia, 2018)

$$\sigma_e = 4\pi R_0^2 \frac{\delta c_0}{\omega_0 R_0} \frac{\Omega^2}{\left(1-\Omega^2\right)^2 + \delta^2 \Omega^2} \qquad (9)$$

where $\Omega = \omega / \omega_0$. The attenuation coefficient $\alpha$ of a bubble suspension can be computed readily by assuming a linear attenuation law, of which the final result is in the form of

$$\alpha = (20\log_{10} e)\frac{1}{2} n\sigma_e \ [\text{dB/m}^2] \qquad (10)$$

where $e$ is the base of the natural logarithm, and $n$ is the total number of microbubbles per unit volume. Since the attenuation coefficient is proportional to the extinction cross-section, $\sigma_e$ is used to represent the attenuation as a matter of convenience.

## III. ATTENUATION DUE TO NONLINEAR OSCILLATION

Due to the employment of the linear solution Eq.(5), the linear attenuation estimated by Eq.(9) is independent of the excitation pressure. This suggests it is incompetence in interpreting the aforementioned nonlinear attenuation phenomenon. Therefore, we may investigate the attenuation phenomenon directly from the full solutions of Eq.(1). Note that the analytical solutions of Eq.(1) are not available, we numerically solve the equation whose radius-time data are directly fed into Eq.(8) to calculate the attenuation curve due to nonlinear oscillations of a UCA.

## IV. RESULTS

The excitation pressure for attenuation measurements in the acoustic experiment usually takes below 100 kPa (about 1 atm) to restrict oscillation of an UCA in the linear region. Here we primarily simulate the dynamic responses of an UCA under 4 different excitation pressures. The shell parameters of the UCA employed in the simulation are radius $R_0 = 2.6\times 10^{-6}$ m, the dilatational viscosity $\kappa^s = 1\times 10^{-9}$ N.s/m, the dilatational elasticity $E^s = 0.04$ N/m, and the reference surface tension $\gamma_0 = 0$. The above values are typical shell parameters of a lipid encapsulated UCA estimated from an attenuation experiment in which an unfocused transducer with a central frequency of 2.25 MHz was used (Xia et al., 2015). The simulated results are listed in **Figure 1**, of which the first row is the instantaneous bubble radius and the corresponding power spectrum in the frequency domain that is converted by fast Fourier transform at the excitation pressure of 0.1 atm. **Figure 1**(a) displays a regular sinusoidal curve with a maximum radius less than 1% of the initial radius. At such a low excitation pressure, the UCA is considered oscillating linearly. This can be confirmed by check the fundamental response in the frequency domain, which is presented in **Figure 1**(b). The normalized power at the fundamental frequency $f/f_e = 1$ (here $f_e$ stands for the excitation frequency) is about 100%, indicating that no energy transfers to other frequency components, and thus the UCA oscillates linearly, completely. This is an obvious result as the excitation pressure is 0.1 atm (about 10 kPa). When the excitation pressures are increased to 0.5 atm and 1 atm, the radius-time curve in **Figure 1**(c) and (e) still looks regular, and their corresponding fundamental responses are about 99.95% and 99.81%, as displayed in **Figure 1**(d) and (f), respectively. Therefore, it is still legitimate to assume that the UCA oscillates linearly at the excitation pressure as high as 1 atm.

When the excitation pressure is further increased to 1.5 atm, **Figure 1**(g) clearly shows the irregular radius-time curve with a maximum radius more than 30% of the initial radius, and the fundamental response in **Figure 1**(e) is only 27.27%. A large part of energy transfers to the subharmonic frequency ($f/f_e = 1/2$).

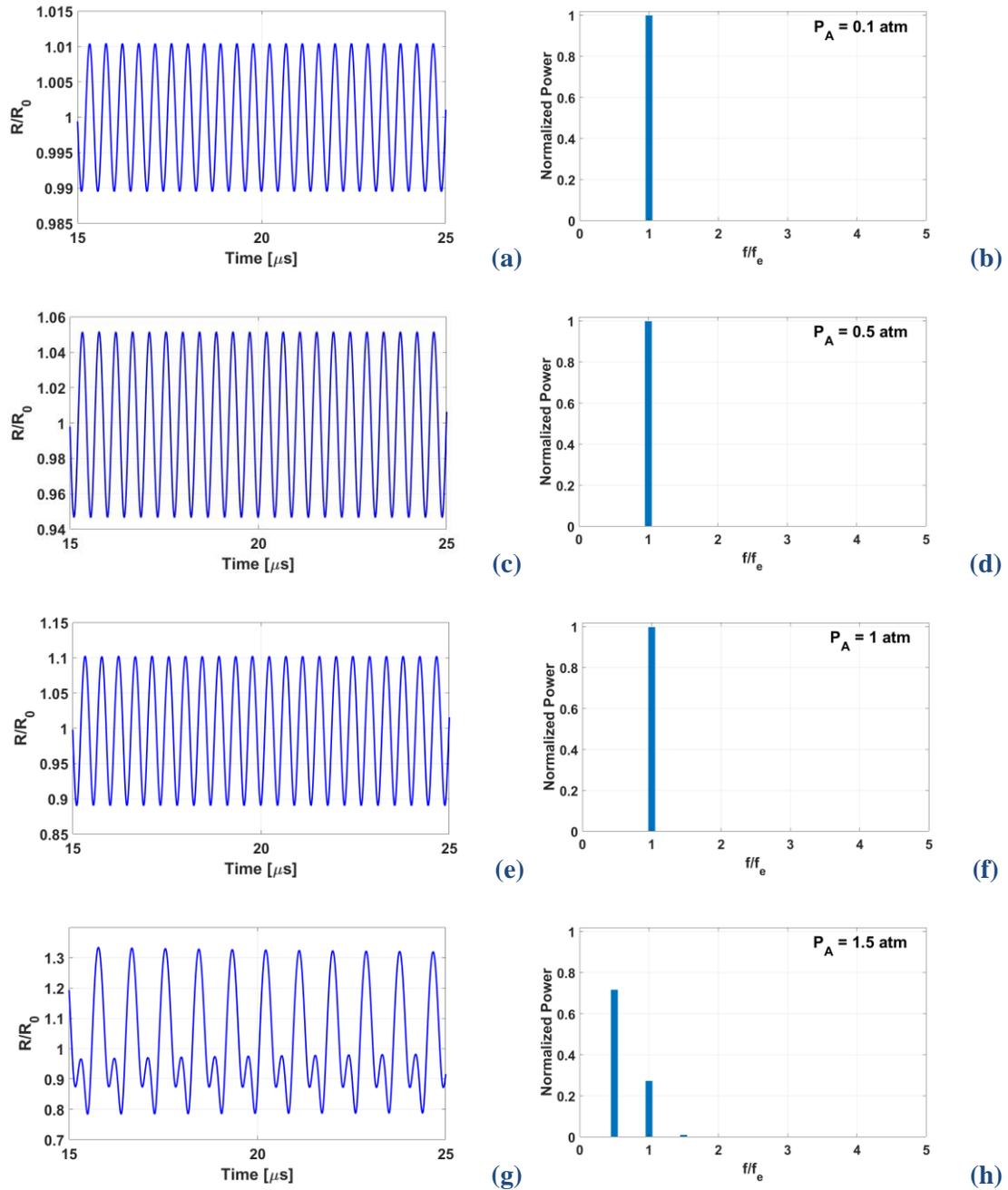

**Figure 1**: The instantaneous bubble radius at the excitation pressure of 0.1 atm (a) and the corresponding power spectrum in the frequency domain (b). (c) and (d), (e) and (f), (g) and (h) are iterated computation with the excitation pressures being increased to 0.5, 1, and 1.5 atm, respectively.

To better visualize the dynamic behavior of the UCA, we plot the dependence of the fundamental response with respect to the excitation pressure in **Figure 2**. The fundamental response is above 99.50% and almost constant before the excitation pressure reaches to 1.2 atm. For this particular UCA, interestingly, the fundamental response shows an inversed 'threshold'. The nonlinear oscillation occurs as the excitation pressure goes beyond 1.2 atm. Therefore, to guarantee linear oscillation of the UCA, we may conclude that the excitation pressure should be less than 1 atm.

However, does the linear oscillation of an UCA also guarantee the linear attenuation of the UCA in the acoustic field? To answer this question, we compare the attenuation coefficient calculated from the linear attenuation theory [Medwin's formula Eq.(9)] and from the nonlinear attenuation formula Eq.(8). *Figure 3*(a) shows that the attenuation curves estimated by the above two formulae are completely coincident at the excitation pressure of 0.1 atm. Due to the resonant effect of the UCA, attenuation curves will display a peak at the main resonance frequency $f_0$. In this figure, the single peak locating at $f/f_0 = 1$ suggests that the UCA oscillates linearly, which was seen in **Figure 1**(a). Where the excitation pressure is increased to 0.5 atm, however, the coincidence of the attenuation curves is broken. The Medwin's estimation (solid blue curve) in *Figure 3*(b) does not change, compared with that of *Figure 3*(a), whereas the attenuation estimated by the full solution Eq.(8) (dashed red curve) displays a skew of the resonance peak and a small tip at $f/f_0 = 0.5$. The excitation frequency at the half of the resonance frequency of the UCA induces the second resonance at the second harmonic, and thus the tip occurs. This indicates nonlinear oscillation happens at the excitation pressure of 0.5 atm. Here we denote the nonlinear attenuation as the deviation of the curve calculated by the full solution from the Medwin's calculation. Note that the same UCA has shown almost linear dynamic behaviors in **Figure 1**(c) and (d). This 'discrepancy' is due to the differences in the excitation frequency that are used in the simulations. The radius-time curves were computed with an excitation frequency of 2.25 MHz, while the attenuation curve estimated by using all the frequencies ranging from 0 to $2f_0$. However, if we look at the attenuation curves in *Figure 3*(b), we can easily find that the curves overlap with each other at the frequency $f/f_0 > 1.5$. Note that the resonance frequency of the UCA is 1.21 MHz, and thus the attenuation is still linear for that range.

The above analysis has a practical significance: UCAs in a suspension are usually sonicated by a broadband ultrasonic transducer with a central frequency, e.g., 2.25 MHz. Even an UCA oscillates linearly at the central frequency, it could still exhibit nonlinear behaviors at other harmonic frequencies due to resonant effects.

*Figure 3*(c) and (d) show more nonlinear phenomena as the excitation pressures increase from 0.5 atm to 1 atm and 1.5 atm. Firstly, the attenuation peak at the resonance of the UCA ($f/f_0 = 1$) skews to a lower frequency of $f/f_0 = 0.84$ in *Figure 3*(c) and $f/f_0 = 0.76$ in *Figure 3*(d). This observation is surprisingly similar to the pressure-dependent attenuation phenomenon that was observed in acoustic experiments (Xia et al., 2015). Secondly, more attenuation peaks occur at $f/f_0 < 1$. This is because the excitation frequency approaches to the second/third resonance frequencies of the UCA. Lastly, the magnitude of the peak attenuation at the resonance decrease from 58.13 in *Figure 3*(b), to 55.66 in *Figure 3*(c) and 52.18 in *Figure 3*(d). This is due to the nonlinear oscillation of the UCA under ultrasound excitation, and the energy at the main resonance frequency transfers to other harmonic components. However, another experimentally observed nonlinear attenuation phenomenon that the magnitude of the attenuation peak at main resonance increases with increasing the excitation pressure is not seen in the present simulation. This could indicate the existing of nonlinear shell behavior of the UCA (Xia et al., 2015).

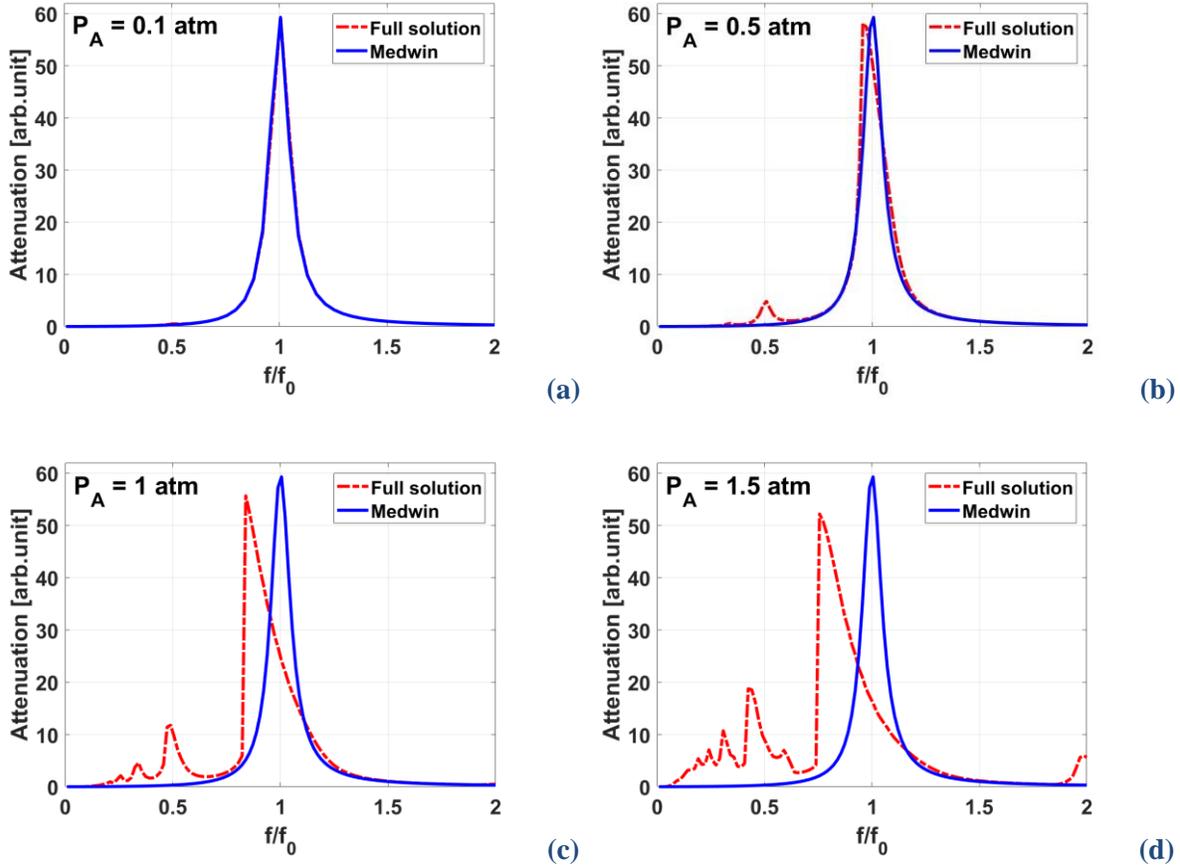

**Figure 3**: The frequency dependent attenuation curves estimated from the full solution Eq.(8) (dashed red) and Medwin's formula (solid blue) at different excitation pressures (a) 0.1 atm, (b) 0.5 atm, (c) 1 atm, and (d) 1.5 atm.

## V. CONCLUSION

By gradually increasing the excitation pressure, we analyzed the dynamic behaviors of an oscillating UCA and calculated the corresponding variations of the attenuation coefficient that was represented by the extinction cross-section. Simulation results for a lipid-encapsulated UCA show the attenuation phenomenon can be nonlinear at the excitation pressure below 1 atm, regardless of the almost linear oscillation of the UCA. These nonlinear behaviors include skewing the main resonance peak to a lower frequency, decreased attenuation at the main resonance, and occurrences of attenuation peaks at sub-harmonics and ultra-harmonics, the first nonlinear phenomenon of which shows great similarities to that observed in experiments. This simulation indicates that the nonlinear dynamics of the UCA at low excitation pressures does affect the accuracy of the attenuation coefficient estimated from the linear attenuation theory Eq.(9). Although the pressure-dependent attenuation phenomena are clearly observed in the simulations, we were not able to reproduce an experimental observation that the attenuation peak at the main resonance increases with increasing the excitation pressure. Therefore, the results in this note suggest that the attenuation phenomenon due to UCAs is rather complicated, and the dynamic behaviors and shell properties of the UCAs could conspire to trigger the nonlinear attenuation behaviors.